# Information Based Centralization of Locomotion in Animals and Robots

Izaak D. Neveln[a,1], Amoolya Tirumalai[a], and Simon Sponberg[a,b]

[a]School of Physics, Georgia Institute of Technology; [b]School of Biology, Georgia Institute of Technology

**Movement in biology is often achieved with distributed control of coupled subcomponents, e.g. muscles and limbs. Coupling could range from weak and local, i.e. decentralized, to strong and global, i.e. centralized. We developed a model-free measure of centralization that compares information shared between control signals and both global and local states. A second measure, called co-information, quantifies the net redundant information the control signal shares with both states. We first validate our measures through simulations of coupled oscillators and show that it successfully reconstructs the shift from low to high coupling strengths. We then measure centralization in freely running cockroaches. Surprisingly, extensor muscle activity in the middle leg is more informative of movements of all legs combined than the movements of that particular leg. Cockroach centralization successfully recapitulates a specific model of a strongly coupled oscillator network previously used to model cockroach leg kinematics. When segregated by stride frequency, slower cockroach strides exhibit more shared information per stride between control and output states than faster strides, indicative of an information bandwidth limitation. However, centralization remains consistent between the two groups. We then used a robotic model to show that centralization can be affected by mechanical coupling independent of neural coupling. The mechanically coupled bounding gait is decentralized and becomes more decentralized as mechanical coupling decreases while internal parameters of control remain constant. The results of these systems span a design space of centralization and co-information that can be used to test biological hypotheses and advise the design of robotic control.**

Motor Control | Information Theory | Locomotion |

A nimal locomotion, the task of actively moving from one position and orientation to another, is achieved through complex dynamics where control is typically distributed across many actuators. For effective locomotion, coordination of muscles and limbs in space and time is necessary to produce directed forces. Locomotor coordination could either be achieved through strong, global coupling with dense connections between components, or though weak, local coupling with sparse connections (1). The continuum between these extreme coupling paradigms is thought of as the centralization/decentralization axis of locomotor control, though a quantifiable measure that can be applied across various systems is lacking. For example, Brambilla *et. al.* defines a decentralized robotic swarm to consist of autonomous individuals that communicate locally and receive no global information (2). Cruse *et. al.* define stick insect motor control as decentralized because muscle commands are more modulated by peripheral feedback rather than the central nervous system (3). However, strong mechanical coupling and feedback from global states could also result in a highly centralized control architecture (4). Methods to assess the empirical centralization of locomotor systems, preferably without any assumption of an underlying dynamic model, are necessary to answer questions regarding how a multi-actuated system is coupled through mechanics, feedback, and control as shown in Fig. 1.

One unresolved locomotor hypothesis is that for fast movements, control via sensory feedback is less effective due to inherent time delays (5). This hypothesis predicts a reliance more on fast decentralized mechanical and neural responses local to each leg where global information decreases with speed faster than local information (4). While there is some evidence that neural feedback is too slow to effectively coordinate control for fast locomotion from experiments in cockroaches (6) and flies (7), some examples of fast local feedback exists (8, 9). An alternative hypothesis is that internal feedforward control may need to be highly centralized to maintain dynamic stability at high speeds (10). There is some evidence that overall coupling increases with speed (11) and that precision in timing of leg movements is coordinated through internal neural coupling(7, 10). However, the challenges of measuring centralization in a system, especially without a specific modeling framework, leaves the general questions regarding the varying degree of centralization in control of animal movement largely unresolved.

Many potential model-based measures exist for quantifying the centralization of systems. Given a full network model, node centrality can indicate which points in the network most govern information flow (12). Distributions of node centrality over networks also indicate overall network architecture (13). For locomotion, we are more interested in a centralization measure that encapsulates the dynamics of how networks coordinate. Coupled-oscillator network models can exhibit coordinated or synchronized behavior similar to the coordination of neural networks or the mechanics of limbs in animals (14). The Kuramoto model of many globally coupled oscillators has been well characterized (15) where oscillators transition from endless incoherence to fast synchronization as coupling increases and global influences outweigh local influences (16). The ability to reduce a large network of oscillators to low dimensional coupled phases or a single global phase described by an order parameter while still capturing dynamics of the system, especially under perturbations, could also indicate a highly centralized architecture (17). Coupled oscillators have been used to represent networks of central-pattern generator (CPG) circuits that drive leg movements (14, 18). These coupled-oscillator models have been used to estimate coupling strengths between control of legs in animal systems (11) as well as controllers for robotic systems (19).

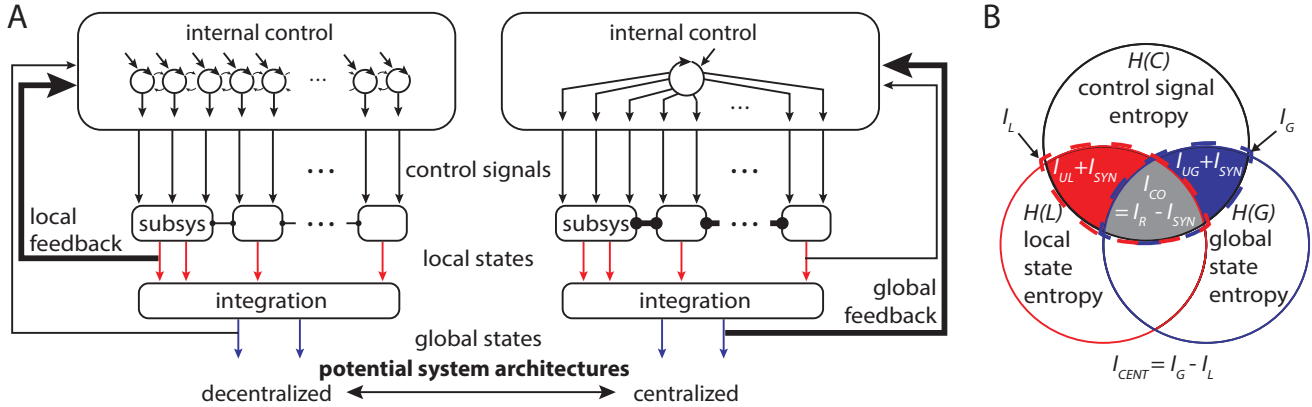

**Fig. 1.** A) Complex dynamical systems can use numerous possible control signals which cascade information through successive levels of integration to effect a relatively small number of global task related outcomes. Coupling between control can range from weak and local (decentralized) to strong and global (centralized). B) Our centralization measure uses empirical observations of control signals, local states, and global states to infer the coupling of control. We estimate the mutual information between the control signal and both the global and local states. Centralization is $I_G$ minus $I_L$, which removes any $I_{CO}$ between the three variables. We expect decentralized systems to contain more $I_L$ and centralized systems to contain more $I_G$.

While increased coupling of CPGs should result in increased centralization, so should increased mechanical coupling and feedback, and all such forms of coupling should be captured in a centralization measure (Fig. 1). Systems can be coordinated solely through mechanical coupling, such as a passive walker (20). Mechanical coupling can also affect feedback circuits that detect changes in force to one leg due to the lift-off of others as has been investigated in stick insects (21). Interlimb coordination, including energy-efficient gait transitions, can be achieved in a quadruped robot solely through local force sensing without any other communication between the leg controllers (22). Force changes can affect the ability to coordinate locomotion as seen in flies (23). Our measure of centralization should reflect how shifts in mechanics can change overall coupling whether through changes in the passive dynamics or feedback circuits that depend on mechanical signals.

Quantifying all of the concepts of centralization so far described rely on a model of the system. Here we develop an empirical measure that can be used to compare the relative centralization of control across different systems and conditions. What unifies concepts of centralization is the amount of global information a control signal shares about the state of the system compared to the amount of local information. We use an information theoretic approach which can assess the dependencies among types and contexts of various locomotor signals. This approach also allows us to separately measure how much net information the control signal shares with both local and global states in a quantity called co-information. We validate these measures of centralization and co-information using a coupled oscillator network of locomotion to ensure that it can reconstruct changes in a model where centralization has been previously defined as the coupling strength. We then analyze the centralization of the coupled oscillator model as it relates to cockroach locomotion and apply our measure to test the hypothesis that cockroach control becomes more centralized at faster running speeds. We also examine how a mechanically coordinated robot is decentralized under varying inertial loadings in order to test if the measure can detect shifts in mechanical coupling. We discuss how these various systems map onto an information space containing centralization and co-information that can be used as a tool for comparing biological control strategies as well as designing robotic control strategies.

## An Information Theoretic Measure of Centralization

Locomotor control is spread out among many control signals that affect local subsystems (Fig, 1A). In the cockroach, control of the muscles in the legs drive the movement of that leg locally. The local states of these subsystems, such as the extension of one leg, each contribute to produce global states, such as the average of all leg extensions. The global states drive the system toward reaching the overall goals, such as how the cockroach's six legs together move its body through the environment. This cascade of signals could consist of multiple layers (e.g. the muscles, joints, legs, etc. of the cockroach) and include feedback from any of those layers (for an investigation into different feedback architectures, see (17)). Furthermore, signals could correlate through coupling between subsystems within a layer, such as mechanical coupling between cockroach legs through the body and ground. By measuring the information shared between a chosen set of control signal, local state, and global state, we aim to infer the structure of the control architecture.

Shared information between variables is quantified by mutual information $I$, which can be depicted as the amount of overlapping entropy as shown in Fig. 1B. A background on information theory is presented in Supplementary Information.

An intuitive decomposition of the mutual information between the control signal and the joint local and global states (depicted by the combination of the red, blue and gray areas in Fig. 1B and Fig. S1) into four separate positive values is given by

$$I(C; \{L, G\}) = I_{UL} + I_{UG} + I_R + I_{SYN}, \quad [1]$$

where $I_{UL}$ and $I_{UG}$ represents information shared uniquely between the control signal and the local and global states respectively, $I_R$ is redundant information shared when either of local or global states are known, and $I_{SYN}$ is synergistic information shared only when both states are known. The axioms that allow for such a decomposition are debated, and estimating these quantities becomes challenging (24). We

2 |

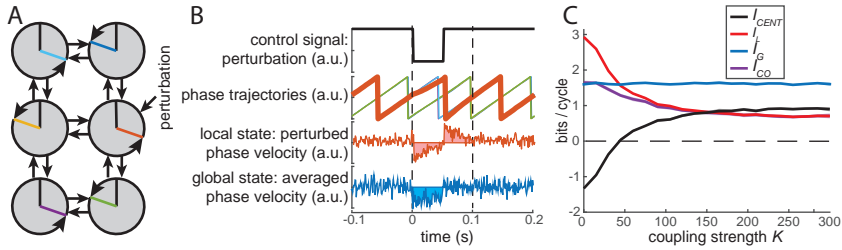

**Fig. 2.** A) We simulated this six phase oscillator network that has been used to model the alternating tripod gait of cockroaches. B) The control signal was a perturbation to a single oscillator with random amplitude. The local state was the integrated absolute deviation from baseline of the phase velocity of the perturbed oscillator, shown by the shaded red region. The global state was similarly calculated from the average phase velocity. C) $I_L$ decreased rapidly with increased coupling strength while $I_G$ remained constant, resulting in a monotonic increase from negative $I_{CENT}$ to positive $I_{CENT}$. $I_{CO}$ coincides with $I_G$ at $K = 0$ and coincides with $I_L$ at high K.

avoid estimating these quantities directly, as we can compute differences of these quantities that are useful measures of the systems we study from our estimates of local MI $I_L$, global MI $I_G$, and total MI $I_{TOT}$ as shown in Fig 1B. For more discussion on the estimation methods used to calculate mutual information, see the Supplementary Material.

We define our measure of centralization $I_{CENT}$ to be

$$I_{CENT} = I_G - I_L = I_{UG} - I_{UL} \quad [2]$$

and thus quantifies the balance between the amount of information about $C$ that is uniquely global versus that which is uniquely local. Redundant or synergistic information does not contribute to centralization.

Co-information $I_{co}$, the gray region in Fig. 1B, is given by

$$I_{CO} = I_L + I_G - I_{TOT} = I_R - I_{SYN}, \quad [3]$$

and it is a measure of net redundancy and does not contain the unique information (24). A negative value indicates that synergistic information outweighs redundant information. $I_{CENT}$ and $I_{CO}$ are therefore two measures that look at how different parts of the total information balance and can both be potentially useful to discriminate different types of neuromechanical control systems.

Grounding these measures back into the specific biological signals, a positive value of $I_{CENT}$ indicates that control signal from the selected muscle is more informative about the global kinematic state averaged from all limbs than the local kinematic state of the leg in which the muscle resides. Also, positive $I_{CENT}$ guarantees non-zero $I_{UG}$, meaning there must be global information not present locally. Therefore, this global information would have to come from some source of coupling (mechanical or neural) within the system. A positive value for $I_{CO}$ indicates some net redundancy between local and global information. As $I_{CO}$ increases, it become less important to know both local and global states to have information about the control signal.

### Results

**Centralization of a Coupled Oscillator Model Increases with Coupling Strength.** We first test whether our measure of centralization captures the previous model-specific definition of centralization based on the strength of a coupled oscillator network shown in Fig. 2A. Not only are coupled oscillator models used as a tool to understand locomotion (14), but this particular model has been previously used to estimate coupling parameters between the six legs used in insect locomotion (11). The dynamics of each oscillator $\theta_i$ is given by

$$\dot{\theta}_i = 2\pi f_i + \sum_{j=1}^{6} K a_{ij} \sin(\theta_j - \theta_i - \phi_{ij}) + 2\pi \nu_i + 2\pi P_i, \quad [4]$$

where $f_i$ is the natural frequency of each oscillator (set to 10 Hz is all cases), $a_{ij}$ is 1 if there exists a connection between oscillator $i$ and $j$ and is zero otherwise, $\phi_{ij}$ is the preferred phase difference between oscillator $i$ and $j$, $\nu_i$ is additive gaussian noise (0 Hz mean, 0.03 Hz standard deviation), and K is the coupling strength between oscillators. We integrate Eqn. 4 using the Euler-Maruyama method (25).

We want to characterize how the information present in a perturbation to an oscillator is spread throughout the network. We prescribe a square pulse $P_i$ lasting one half cycle put into just one oscillator as shown in Fig. 2B, which we use as the control signal $C$, with an amplitude drawn from a random Gaussian distribution ($-5$ Hz mean, $\frac{4}{3}$ Hz standard deviation). We then measure both the local response of that oscillator (the integrated deviation away from the steady state phase velocity) to that input and the average global response (same as local only all phase velocities are averaged together) of all oscillators as shown in Fig. 2B.

As shown in Fig. 2C, the system is fully decentralized when $K = 0$, meaning $I_L$ outweighs $I_G$. $I_{CO}$ matches $I_G$ indicating that any $I_G$ is redundant with local information resulting in no $I_{UG}$. As the perturbation cannot propagate to the other oscillators, no additional information can be present in the global signal. As coupling is introduced and increases, $I_{CENT}$ increases, becomes positive, and levels out to a maximal value. At these high coupling strengths the $I_L$ is completely redundant, meaning the value for $I_{CENT}$ equals the amount of $I_{UG}$. Thus, though $I_G$ stays constant with increased coupling strength, $I_{UG}$ must increase from zero to a positive value as a positive value of $I_{CENT}$ requires there exist $I_{UG}$. Changes to coupling strength can manifest in physical oscillators through changing the mass of a freely moving platform that holds a number of metronomes (26) or increasing the number of connections between central pattern generating circuits driving locomotion (27). Our centralization measure could empirically determine the relative coupling strengths of these systems, validating it as a useful diagnostic tool.

**Cockroach Centralization During Running.** We ran 9 cockroaches over flat terrain while recording EMG activity from the femoral extensor muscle 137 in the middle leg and tracking the 2D kinematics of the ends of all 6 legs as shown in Fig. 3A. This muscle has previously been implicated in control even during high speed running (29). See Supplementary Methods section *Cockroach Experiments* for more details on the experimental protocol.

When analyzing 2343 strides from all cockroaches (Fig. 3B-F), we find that $I_{CENT}$ is positive (Fig. 3G). Positive $I_{CENT}$ means motor unit spikes are more informative about the global average kinematics than the local kinematics of the limb where

Neveln *et al.* April 17, 2019 | 3

the control signal originates. It is surprising that the activation of a muscle in one leg indicates more about the average state of all the limbs than that of the leg in directly activates. This main result is likely because of strong neural and mechanical coupling between the legs (11, 30).

When just spike count is considered, $I_L$ slightly outweighs $I_G$, though the proportion of overall information is small. Most information, and the positive value for $I_{CENT}$, arises only when spike timing is also factored into the analysis. Estimates of mutual information between the control signal and the local and global states were stabilized when just a two-dimensional representation of the states were used (see SI Fig. S2-3 and Supplementary Text).

Recent studies show that the timing of individual motor unit spikes has causal effects on motor dynamics down to the millisecond scale(31, 32). Our results show that most of the dependencies between the control signal and the local and global states only manifest when the timing of the spike relative to the phase of the stride is considered. Many analyses of motor neuronal activity in insects use only the rate of activity (33–35). It is possible that much of the encoded information regarding leg coupling is suppressed in such analyses.

Muscle 137 (as well as its homologous muscle 178 in the hind leg) is driven by a single fast motor neuron that also drives other extensor muscles 136, 135d, and 135e (179, 177d, and 177e in the hind leg). (36). These muscles can produce varying mechanical work from the same signal (37), including positive work to drive extension or negative work to slow flexion. Therefore, this single motor unit has been implicated in the control of leg flexion and reversal (38) and the start of joint extension and stride length (29). Our results for the middle leg indicate the control signal shares non-redundant information with both the stance and swing portions of the stride, which both corroborates the reported versatility of this motor unit as well as the observation that muscle work depends on the state of the limb (39).

**The Effect of Speed on $I_{CENT}$ and $I_{CO}$.** Given that the cockroaches tested exhibited a wide range of stride frequencies (Fig. 3B), we can test whether faster strides were more centralized possibly for maintaining dynamic stability (10) or more decentralized possibly due to bandwidth constraints (4). When we segment the cockroach data into two groups according to stride frequency, we observe $I_{CENT}$ does not change (leftmost column of Fig. 4). However, both $I_G$ and $I_L$ do change in interesting ways.

$I_G$ and $I_L$ per stride is higher for slower strides than for faster strides (Fig. 4A). This difference is due to a similar trend when looking at timing information. $I_G$ and $I_L$ when just spike count is considered is slightly lower in the slower group, though again count information contributes much less information overall. When converted to bits per second using the median frequency of each group (Fig. 4B), we actually see that the information per unit time (bit rate) is greater for the faster group.

Though the balance of local and global information does not change, perhaps the two states become more redundant with greater speed. Overall, $I_{CO}$ is similar between fast and slow groups. However, the faster group closer to full redundancy as the $I_{TOT}$ is smaller. The slower group has higher $I_{CO}$ in timing and lower $I_{CO}$ in count. Timing is therefore more redundant for the slow group, whereas $I_{CO}$ is actually negative in count indicating some degree of synergy between local and global signals. Therefore, for the slow group, the local and global output together are more informative on the number of spikes in a stride than when taken separately.

The amount of information available to be transferred from motor neuron to muscle to leg output and back again through feedback is expected to decrease for faster strides(6), which is what we see in our data. However, the decrease per stride is not as much as expected with the assumption of a constant information rate. The predictions for $I_{CENT}$ are complicated because while this information decrease with speed is expected, internal and mechanical coupling is hypothesized to increase to maintain dynamic stability (40). Spatial coordination (7) and temporal coordination (34, 41) has been shown to increase with speed (7). Spatial coordination degrades when sensory feedback is disrupted (10). When fitting thoracic ganglia burst activity to coupled-oscillator models, no correlation is observed between burst frequency and coupling strength (42) and a very weak positive correlation between running speed and coupling strength was observed when fitting free-running cockroach leg kinematics to such a model (11). Our measure of centralization, which takes into account a neural control signal with local and global kinematics indicates no shifts in overall coupling with speed when considering cockroach running.

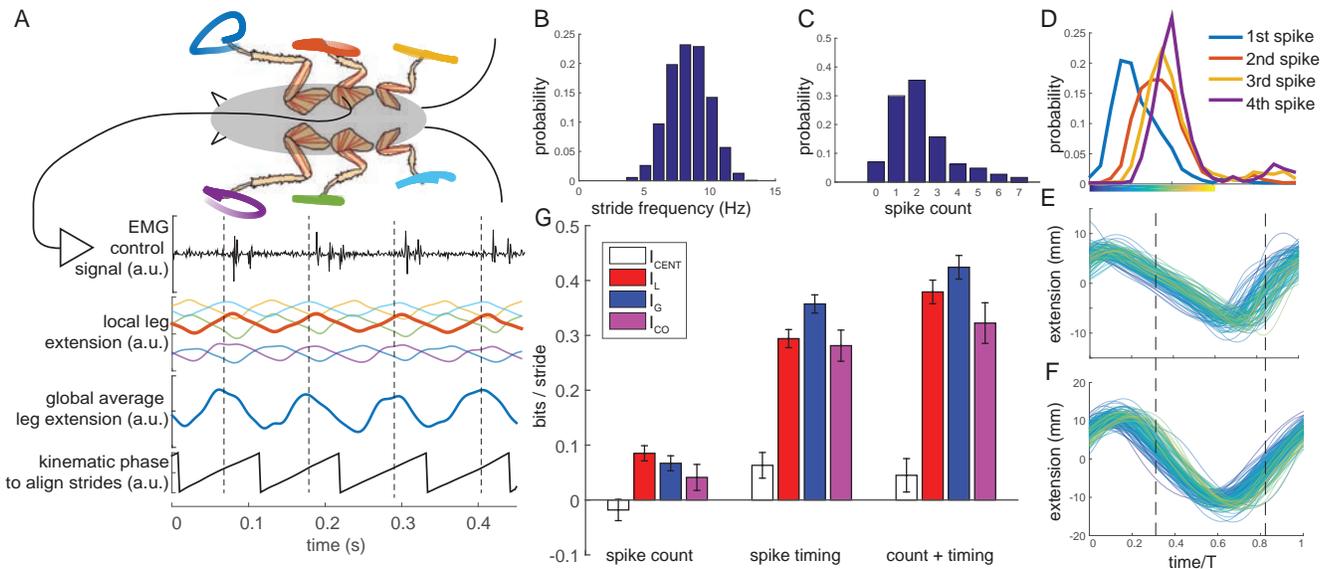

**Fig. 3.** A) Control, local, and global signals recorded from the cockroach. Strides were separated according to kinematic phase calculated using the Phaser algorithm (28). B) Distribution of stride frequencies across all 2343 strides taken from 9 animals. C) Distribution of the number of spikes in the femoral extensor over a stride. D) Probability density functions of the timing of the first four spikes if present, with time normalized by stride period $T$. E) Local leg extension trajectories colored by the timing of the first spike (colormap from D). Correlations between the timing of the first spike and the resulting local and global states are visible in trajectories when the first spike occurred early in the stride colored blue are distinguished from those when the first spike occurred late in the cycle colored yellow. F) Global leg extension trajectories colored as in E. G) $I_{CENT}$, $I_G$, $I_L$, and $I_{CO}$ for all strides.

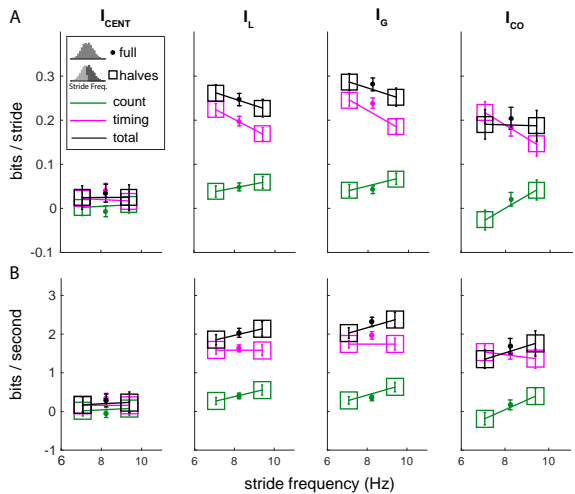

**Fig. 4.** A) Fast versus slow strides. A) When splitting the strides into slow and fast halves, $I_{CENT}$ remains unchanged. However, both local and global mutual information decrease equally, and this decrease is attributed to information in timing. $I_{CO}$ remains constant due to competing trends in timing and count. B) Same as (A) only information is converted to a bits/second rate using the median stride frequency of each group. Though the information per stride decreases, the information per unit time increases.

**Coordination Through Mechanical Coupling is Decentralized in Robot Bounding Gate.** If neural feedback delays are too long to effectively couple limbs during fast locomotion to properly respond to perturbations, mechanical coupling could potentially compensate. Furthermore, changes to mechanical coupling will alter feedback signals related to the state of the system and its parts. Clearly, mechanics must be considered when analyzing the control architecture of dynamic locomotion. The Minitaur robot (Ghost Robotics, Inc. Philadelphia, PA) shown in Fig. 5A demonstrates coordination through mechanical coupling (43). As one pair of legs impacts the ground, the rest of the body translates and rotates, generally resulting in movement of the hips of the alternate leg pair. Therefore, even if the commanded torque to one leg pair does not explicitly depend on the states of the other leg pair (i.e. no internal 'neural' coupling), the two leg pairs will tend towards a bound gait where the front pair of synchronized legs alternates with the synchronized back pair, or a pronk gait where all legs are synchronized. Transitions between these gaits can occur by changing mechanical coupling through changes to the moment of inertia $M$ around the pitch axis, or by adding phase coupling into the internal control block.

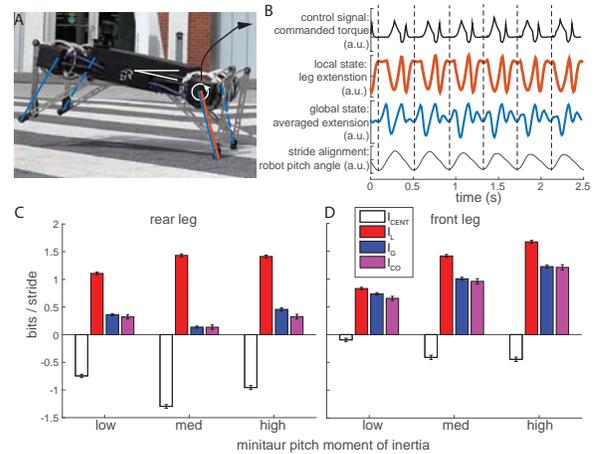

**Fig. 5.** A) Image of Minitaur indicating the variables measured for the centralization calculation. B) The commanded torque at the hip to drive the extension of a single leg was selected as the control signal. The local state was the measured extension of that leg. The global state was the average of all leg extension trajectories. Strides were aligned by the pitch angle of the robot. C) $I_{CENT}$ and $I_{CO}$ of the three different moment of inertia conditions. The torque to a rear leg was used as the control variable. D) Torque to a front leg was used as a control variable.

We altered $M$ by shifting two weights in opposite directions longitudinally along the robot, thus keeping the center of mass constant. These varying conditions test whether our empirical centralization measure can detect changes to mechanical coupling. We also predict $I_{CO}$ will be positive and close to maximal, because any information transfer through mechanical coupling should be redundant if the system is relatively stiff.

We ran the bound gait described in (43) over flat terrain, which still produces variability in each stride, and measured local mutual information between the torque command and the actual extension of that leg as shown in Fig. 5B. We compare the local information to the global mutual information between that same torque signal and the average extension trajectories of all four legs. As shown in Fig. 5, we compare low, intermediate, and high values of $M$.

$I_L$ is greater than $I_G$ for all experimental conditions, resulting in a negative value of $I_{CENT}$. For the more decentralized rear leg pair, $I_{CENT}$ is minimized for the intermediate $M$ condition, confirming the prediction for when mechanical coupling is minimized. $I_{CENT}$ is greatest for the low $M$, where $I_G$ is fully redundant. For the front leg pair, $I_{CENT}$ is minimized for both the intermediate and high $M$ conditions.

$I_{CENT}$ of control for signals from the front legs is overall higher when considering the front leg versus the back leg, indicating an asymmetry to the mechanical coupling not predicted in the reduced models that only consider bounding in place. As this asymmetry is the same regardless of which of the front or back legs are analyzed, we expect that this difference is partly due the forward movement of the robot. This result is an example of how measuring $I_{CENT}$ can result in new discoveries that may not be predicted from simplified models.

Consideration of mechanics is necessary for understanding locomotor control (30). The virtual leg of running animals all use a similar non-dimensionalized stiffness that also optimizes locomotion in robots (18), allowing a six legged robot with correctly tuned mechanics to move with just a single actuator (44). Adding stiff spines to legs (45), flexible joints to the body (46), or streamlined shells (47) allow animals and robots to traverse challenging terrain. The ability to estimate the effects of mechanical feedback such as in these examples could allow for adaptive control (48). Our centralization measurements resolve changes to the mechanics in the robot that would not be evident from kinematics or footfall patterns alone.

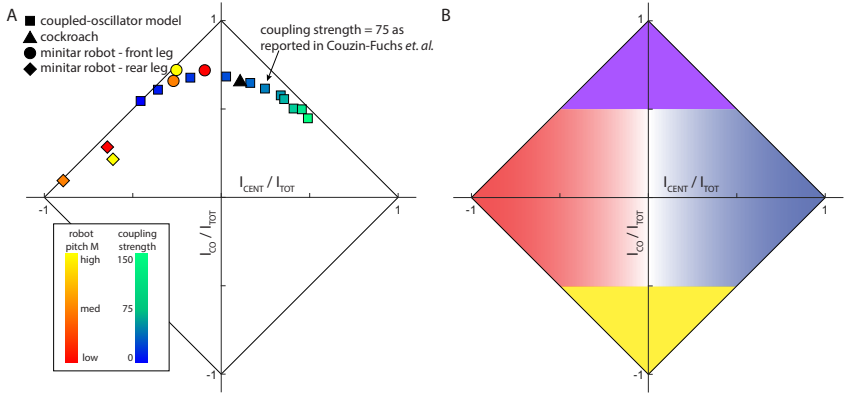

**Fig. 6.** A) Centralization and co-information for all systems normalized by total information. B) The centralization/co-information control architecture space. Any system will fall within a diamond region bounded by the solid black diagonal lines. Centralized systems, where a control signal is more informative about global states than local states, will fall into the blue region. Conversely, decentralized systems will be more weighted toward local information and fall in the red region. The purple region represents systems with high redundancy, where local and global states carry overlapping information about the control signal. The yellow region represents synergistic systems, where knowing both local and global states together with give more information about the control signal than separately.

## Discussion

We have introduced an empirical measure of centralization of locomotor control that does not rely on any underlying model. Therefore, we can use $I_{CENT}$ to compare systems. We choose a normalization scheme that compares $I_{CENT}$ and $I_{CO}$ in proportion to $I_{TOT}$ shown in Fig. 6.

The coupled oscillator model, which has been used to describe legged locomotion (14) and the control of robots (19), has been used previously to represent gradations of control along the centralized/decentralized axis (4). When plotted on the $I_{CENT}/$ - $I_{CO}$ axes in Fig. 6A, the coupled oscillator does vary from decentralized to centralized as coupling strength increases. However, we also see that fast cockroach locomotion is centralized, meaning that $I_G$ outweighs $I_L$, and $I_{CENT}$ increases with stride frequency. The overall $I_{CENT}$ of the cockroach matches that of the coupled oscillator model with a slightly centralized coupling strength. This result is further validated by a previous study which fit cockroach leg kinematics explicitly to a coupled-oscillator model (11). The coupling strengths estimated from this fitting averaged to 76.1, and when we estimate $I_{CENT}$ for this model we find that $I_G$ does indeed outweigh $I_L$ and is very close to the measured $I_{CENT}$ of the cockroach (Fig. 6).

The robotic results contrast that of the cockroach as they fall on the decentralized side of the axis. More importantly, the differences in $I_{CENT}$ due to changes in mechanical coupling show the importance of considering interactions between the limbs, body, and environment when designing control for robotic coordination. Our centralization measure can thus be used as an empirical diagnostic tool to assess how coupling between legs is affected by mechanical changes that could be difficult to model.

The animal, robotic, and coupled oscillator systems explicitly explore the centralization axis. Other systems likely populate the rest of the information space and exploring benefits of the different regions could be a guide in developing robot control or analyzing animal locomotion. While we found no overall change in centralization with speed during running in cockroaches, the constituent informations, $I_L$ and $I_G$, did change (Fig. 6). Changing gait is likely to shift location in the architecture space. Slower walking gaits in cockroaches (49), stick insects (14), and robots (50) are thought to be more more decentralized using local or neighbor-based reflex rules (3). Even though central pattern generator circuits are still involved they are distributed and typically weakly coupled (51) predicting a more decentralized information strategy. Different environments might also demand different information strategies. Tests in robotic models indicate that some amount of decoupling between legs, rather than a single centralized controlled trajectory for all legs, results in increased robustness over more variable terrain (52). These results would predict a leftward shift along $I_{CENT}$ on rough terrain. Movement in either direction on the centralization axis could simplify control, such as a highly actuated ribbon fin that only needs to shift trajectories of several of a hundred fin rays to maneuver (53) might be more decentralized, while few control signals driving the coordination of many muscles might be more centralized (54). Overall, scenarios where positive $I_{CENT}$ is beneficial suggest that it is more important sense the global state whereas scenarios where negative $I_{CENT}$ is beneficial suggest emphasizing local state sensing.

Due to the general collapse of high dimensional control inputs to a low dimensional outputs, one might expect the positive $I_{CO}$ indicating net redundancy between global and local information in all the systems (55). In most all examples shown here, $I_{CO}$ is positive except for the slower group of cockroach data when considering only spike count (Fig. 4). Changes along the $I_{CO}$ axis are possible and could give different performance benefits. For the example of mechanical coupling, stiffer legs and body coupling would likely result in a highly redundant system in the purple area in Fig. 6B. In terms of maximizing the possible information the control variable could share with both the local and global states, it would be beneficial to have synergistic information rather than redundant information. Such a scenario could be possible if the control receives both global and local feedback, where both types of feedback affect the control signal differently together than they do separately. We expect soft animals and robots could benefit from a synergistic control architecture because local states might be very independent from global states. One example is a robotic slime mold where each actuator on the edge receives feedback relating to its local neighbors as well as the inner protoplasm that globally interacts with all actuators (56). The controller takes advantage of the different information the global and local state provides, which would suggest synergy and predict a location in the bottom half of the control architecture spaces.

**ACKNOWLEDGMENTS.** We thank Avik De and Dan Koditschek for facilitating experiments with Minitaur. Shai Revzen, Sam Burden, Ilya Nemenman, Bob Full, Adrienne Fairhall and Sara Solla provided helpful discussions. This work was supported by NSF CAREER MPS/PoLS 1554790 to S.S.

1. Aoi S, Manoonpong P, Ambe Y, Matsuno F, Wörgötter F (2017) Adaptive Control Strategies for Interlimb Coordination in Legged Robots: A Review. *Frontiers in neurorobotics* 11:39.
2. Brambilla M, Ferrante E, Birattari M, Dorigo M (2013) Swarm robotics: a review from the swarm engineering perspective. *Swarm Intelligence* 7(1):1–41.
3. Cruse H, Durr V, Schmitz J (2007) Insect walking is based on a decentralized architecture revealing a simple and robust controller. *Philosophical Transactions of the Royal Society A: Mathematical, Physical and Engineering Sciences* 365(1850):221–250.
4. Holmes P, Full RJ, Koditschek D, Guckenheimer J (2006) The Dynamics of Legged Locomotion: Models, Analyses, and Challenges. *SIAM Review* 48(2):207–304.
5. Brown IE, Loeb GE (2000) A Reductionist Approach to Creating and Using Neuromusculoskeletal Models in *Biomechanics and Neural Control of Posture and Movement*. (Springer New York, New York, NY), pp. 148–163.
6. Sponberg S, Full RJ (2008) Neuromechanical response of musculo-skeletal structures in cockroaches during rapid running on rough terrain. *The Journal of experimental biology* 211(Pt 3):433–46.
7. Mendes CS, Bartos I, Akay T, Márka S, Mann RS (2013) Quantification of gait parameters in freely walking wild type and sensory deprived Drosophila melanogaster. *eLife* 2:e00231.
8. Fayyazuddin A, Dickinson MH (1996) Haltere afferents provide direct, electrotonic input to a steering motor neuron in the blowfly, Calliphora. *The Journal of neuroscience* 16(16):5225–32.
9. Höltje M, Hustert R (2003) Rapid mechano-sensory pathways code leg impact and elicit very rapid reflexes in insects. *The Journal of experimental biology* 206(Pt 16):2715–24.
10. Couzin-Fuchs E, Gal O, Holmes P, Ayali A (2015) Differential control of temporal and spatial aspects of cockroach leg coordination. *Journal of Insect Physiology*.
11. Couzin-Fuchs E, Kiemel T, Gal O, Ayali A, Holmes P (2015) Intersegmental coupling and recovery from perturbations in freely running cockroaches. *Journal of Experimental Biology*.
12. Opsahl T, Agneessens F, Skvoretz J (2010) Node centrality in weighted networks: Generalizing degree and shortest paths. *Social Networks* 32(3):245–251.
13. Watts DJ, Strogatz SH (1998) Collective dynamics of 'small-world' networks. *Nature* 393(6684):440–442.
14. Ayali A, et al. (2015) The comparative investigation of the stick insect and cockroach models in the study of insect locomotion. *Current Opinion in Insect Science* 12:1–10.
15. Kuramoto Y, Nishikawa I (1987) Statistical macrodynamics of large dynamical systems. Case of a phase transition in oscillator communities. *Journal of Statistical Physics* 49(3-4):569–605.
16. Strogatz SH (2000) From Kuramoto to Crawford: exploring the onset of synchronization in populations of coupled oscillators. *Physica D: Nonlinear Phenomena* 143(1-4):1–20.
17. Revzen S, Koditschek DE, Full RJ (2009) Towards testable neuromechanical control architectures for running in *Progress in Motor Control*. (Springer), pp. 25–55.
18. Koditschek DE, Full RJ, Buehler M (2004) Mechanical aspects of legged locomotion control. *Arthropod Structure & Development* 33(3Portland, OregonSeattle, Washington):251–272.
19. Ijspeert AJ, Crespi A, Ryczko D, Cabelguen JM (2007) From swimming to walking with a salamander robot driven by a spinal cord model. *Science* 315(5817):1416–20.
20. McGeer T, , et al. (1990) Passive dynamic walking. *I. J. Robotic Res.* 9(2):62–82.
21. Dallmann CJ, Hoinville T, Dürr V, Schmitz J (2017) A load-based mechanism for inter-leg coordination in insects. *Proceedings. Biological sciences* 284(1868).
22. Owaki D, Ishiguro A (2017) A Quadruped Robot Exhibiting Spontaneous Gait Transitions from Walking to Trotting to Galloping. *Scientific Reports* 7(1):277.
23. Mendes CS, Rajendren SV, Bartos I, Márka S, Mann RS (2014) Kinematic Responses to Changes in Walking Orientation and Gravitational Load in Drosophila melanogaster. *PLoS ONE* 9(10):e109204.
24. Ince R, Ince, A. RA (2017) Measuring Multivariate Redundant Information with Pointwise Common Change in Surprisal. *Entropy* 19(7):318.
25. Kloeden PE, Platen E (1992) *Numerical {S}olution of {S}tochastic {D}ifferential {E}quations*. (Springer Berlin Heidelberg), p. 636.
26. Pantaleone J (2002) Synchronization of metronomes. *American Journal of Physics* 70(10):992–1000.
27. Ijspeert AJ (2008) Central pattern generators for locomotion control in animals and robots: A review. *Neural Networks* 21(4):642–653.
28. Revzen S, Guckenheimer JM (2008) Estimating the phase of synchronized oscillators. *Physical Review E* 78(5):051907.
29. Watson JT, Ritzmann RE (1998) Leg kinematics and muscle activity during treadmill running in the cockroach, blaberus discoidalis: Ii. fast running. *J. Comp. Physiol. A.* 182(1):23–33.
30. Jindrich DL, Full RJ (2002) Dynamic stabilization of rapid hexapedal locomotion. *The Journal of experimental biology* 205(Pt 18):2803–23.
31. Sponberg S, Daniel TL (2012) Abdicating power for control: a precision timing strategy to modulate function of flight power muscles. *Proceedings. Biological sciences* 279(1744):3958–66.
32. Srivastava KH, et al. (2017) Motor control by precisely timed spike patterns. *Proceedings of the National Academy of Sciences of the United States of America* 114(5):1171–1176.
33. Fuchs E, Holmes P, Kiemel T, Ayali A (2011) Intersegmental coordination of cockroach locomotion: adaptive control of centrally coupled pattern generator circuits. *Frontiers in Neural Circuits* 4:125.
34. Fuchs E, Holmes P, David I, Ayali A (2012) Proprioceptive feedback reinforces centrally generated stepping patterns in the cockroach. *Journal of Experimental Biology* 215(11):1884–1891.
35. Mantziaris C, et al. (2017) Intra- and intersegmental influences among central pattern generating networks in the walking system of the stick insect. *Journal of Neurophysiology* 118(4):2296–2310.
36. Pearson KG, Iles JF (1971) Innervation of coxal depressor muscles in the cockroach, Periplaneta americana. *The Journal of experimental biology* 54(1):215–232.
37. Ahn AN, Meijer K, Full RJ (2006) In situ muscle power differs without varying in vitro mechanical properties in two insect leg muscles innervated by the same motor neuron. *The Journal of experimental biology* 209(Pt 17):3370–82.
38. Full R, Stokes D, A (1998) Energy absorption during running by leg muscles in a cockroach. *The Journal of experimental biology* 201 (Pt 7)(7):997–1012.
39. Sponberg S, Libby T, Mullens CH, Full RJ (2011) Shifts in a single muscle's control potential of body dynamics are determined by mechanical feedback. *Philosophical Transactions of the Royal Society of London. Series B, Biological sciences* 366(1570):1606–20.
40. Kukillaya R, Proctor J, Holmes P (2009) Neuromechanical models for insect locomotion: Stability, maneuverability, and proprioceptive feedback. *Chaos* 19(2):026107.
41. Wosnitza A, Bockemühl T, Dübbert M, Scholz H, Büschges A (2013) Inter-leg coordination in the control of walking speed in Drosophila. *The Journal of experimental biology* 216(Pt 3):480–91.
42. David I, Holmes P, Ayali A (2016) Endogenous rhythm and pattern-generating circuit interactions in cockroach motor centres. *Biology Open* 5(9).
43. De A, Koditschek DE (2018) Vertical hopper compositions for preflexive and feedback-stabilized quadrupedal bounding, pacing, pronking, and trotting. *The International Journal of Robotics Research* 37(7):743–778.
44. Hoover AM, Burden S, Xiao-Yu Fu, Shankar Sastry S, Fearing RS (2010) Bio-inspired design and dynamic maneuverability of a minimally actuated six-legged robot in *2010 3rd IEEE RAS & EMBS International Conference on Biomedical Robotics and Biomechatronics*. (IEEE), pp. 869–876.
45. Spagna JC, Goldman DI, Lin PC, Koditschek DE, Full RJ (2007) Distributed mechanical feedback in arthropods and robots simplifies control of rapid running on challenging terrain. *Bioinspiration & Biomimetics* 2(1):9–18.
46. Jayaram K, Full RJ (2016) Cockroaches traverse crevices, crawl rapidly in confined spaces, and inspire a soft, legged robot. *Proceedings of the National Academy of Sciences of the United States of America* 113(8):E950–7.
47. Li C, et al. (2015) Terradynamically streamlined shapes in animals and robots enhance traversability through densely cluttered terrain. *Bioinspiration & Biomimetics* 10(4):046003.
48. Sastry SS, Bodson M (2011) *{A}daptive {C}ontrol: {S}tability, {C}onvergence and {R}obustness*. (Courier Corporation).
49. Bender JA, et al. (2011) Kinematic and behavioral evidence for a distinction between trotting and ambling gaits in the cockroach Blaberus discoidalis. *The Journal of experimental biology* 214(Pt 12):2057–64.
50. Schilling M, Hoinville T, Schmitz J, Cruse H (2013) Walknet, a bio-inspired controller for hexapod walking. *Biological Cybernetics* 107(4).
51. Büschges A, Borgmann A (2013) Network Modularity: Back to the Future in Motor Control. *Current Biology* 23(20):R936–R938.
52. Von Twickel A, Büschges A, Pasemann F (2011) Deriving neural network controllers from neuro-biological data: Implementation of a single-leg stick insect controller. *Biological Cybernetics* 104(1-2):95–119.
53. Sefati S, et al. (2013) Mutually opposing forces during locomotion can eliminate the tradeoff between maneuverability and stability. *Proceedings of the National Academy of Sciences of the United States of America* 110(47):18798–803.
54. Ting LH, Macpherson JM (2005) A Limited Set of Muscle Synergies for Force Control During a Postural Task. *Journal of Neurophysiology* 93(1):609–613.
55. Full RJ, Koditschek DE, Full RJ (1999) Templates and anchors: neuromechanical hypotheses of legged locomotion on land. *The Journal of Experimental Biology* 2(12):3–125.
56. Umedachi T, Takeda K, Nakagaki T, Kobayashi R, Ishiguro A (2010) Fully decentralized control of a soft-bodied robot inspired by true slime mold. *Biological Cybernetics* 102(3):261–269.

**Supporting Information Text**

**Background on Information Theory.** The discrete Shannon entropy ($H$) of a signal, given by the equation

$$H(S) = \sum_i p(s_i)\log p(s_i), \quad [1]$$

quantifies the amount of information present in the signal, where $s_i$ is each possible state the signal $S$ can take and $p$ is the probability distribution of the states (1). When the base of the logarithm is two, the unit of entropy is bits, where the number of bits represents the expected number of yes or no questions to determine the state of the signal. Entropy can be similarly defined for joint distributions as $H(S_1; S_2)$ and conditional distributions as $H(S_1|S_2)$.

$$H(S_1; S_2) \leq H(S_1) + H(S_2) \quad [2]$$

with equality only when the two signals are independent. The level of dependency between two signals is quantified by the mutual information $I$, which is the difference from equality in Eq. 2 given by the equation

$$I(S_1; S_2) = H(S_1) + H(S_2) - H(S_1, S_2). \quad [3]$$

Mutual information can also be written as

$$I = H(S_1) - H(S_1|S_2) = H(S_2) - H(S_2|S_1). \quad [4]$$

Therefore, the mutual information measures the decrease in entropy in one signal when the state of the other signal is known. These overlapping entropies for our chosen set of signals (hereafter labeled $C$ for the set of possible $c_i$ control states, $L$ for the set of local states, and $G$ for the set of global states) are graphically presented in Fig. S1. Estimation of mutual information of continuous variables can have error or bias due to limited sampling (2). We use a bin-less nearest neighbor estimator of $I$ (2) which handles these issues well, as described in the next section.

The entropy diagram in Fig. S1 helps build intuition about how the different mutual information quantities contain different parts of the decomposition of the total mutual information between the control signal and the joint local and global states. Local MI outlined by the dashed red line is the red and gray areas together in Fig. S1 and is given by

$$I_L = I_{UL} + I_R. \quad [5]$$

This is the mutual information between $C$ and $L$ when $G$ is not known. When $G$ is known, then the red area in Fig. S1 is given by

$$I(C; L|G) = I_{UL} + I_{SYN} \quad [6]$$

and does not include $I_R$. Therefore, the grey area must have $I_R$ and a negative $I_{SYN}$ to balance out the positive $I_{SYN}$ in the red area to have zero overall synergy in $I_L$.

**Estimating Mutual Information.** We used the $k$-nearest neighbor method for estimating MI (2). In brief, the underlying conditional probability densities are estimated by counting how many samples in the marginal spaces are contained within the distance to the $k$-nearest neighbor of each sample point in the joint space. The choice of $k$ sets the resolution to which the probability densities are estimated as the method assumes a uniform distribution in the ball smaller than the distance to the $k$-nearest neighbor. For details of the estimator see (2).

We renormalized our variables to have zero mean and unit variance. Such a reparametrization has no impact on the actual MI between two variables, but can produce a better estimate as each variable is scaled equally and outliers have a smaller influence (2). Also, as our spiking variable is discrete, we added a small amount of noise with a standard deviation of $10^{-4}$ as otherwise many points in the dataset would have the same coordinates and therefore counting to the $k$-nearest neighbor becomes impractical.

To calculate the co-information, we first estimated the total mutual information between the control variable and the joint local and global variables. Co-information is given by

$$I_{CO} = I_G + I_L - I_{TOT} \quad [7]$$

We chose a value of $k$ for which the estimates of the different mutual information values remained consistent as $k$ varied. From Fig. S2, values of $k$ between 5 and 10 give the same estimates for count (top plot), and they fall off consistently for timing (bottom plot). These consistent trends mean that the local and global estimates give consistent values for centralization whether or not normalized by the total information. We therefore use a value of $k = 7$ for calculating centralization and note that conclusions do not depend on changing $k$ between 5 and 10.

We followed a procedure similar to that in (3) to determine the error of our estimate of MI. We subsampled the data into $m$ equally sized and independent groups containing $N/m$ samples, calculated the MI for those $m$ groups, and then calculated the standard deviation of those $m$ MI estimates. We repeated this process 10 times and averaged the standard deviations ($\sigma$) for each value of $m$. We fit these mean standard deviations to $\log \sigma^2 = A + \log m$ relationship and estimated $\sigma$ for the original full dataset by setting $m = 1$. The errorbars displayed in Fig. S2 show these measured and extrapolated $\sigma$ values. We are also able to assess whether there exists sample-sized bias in our MI estimates if the estimates of MI stay within the errorbars as $m$ is increased and the sample size is decreased. As shown in the right column of plots in Fig. S2, estimates for count (top plot) remain consistent and estimates for timing (bottom) fall off with the number of groups at the same rate, resulting in similar estimates of centralization whether or not normalized by total information.

**Cockroach Experiments.** *Blaberus discoidalis* (henceforth cockroaches) were kept in an incubation chamber set to 37° C, 60% humidity, and a 12h/12h light cycle with ample supply to food and water. Cockroaches were first cold anesthetized in a refrigerator at 4° C for about 30 minutes. We then removed their wings and cut back their pronotum so that their legs would be more visible for our overhead video recordings.

To insert the electromyogram (EMG) wires, we first restrained them ventral side up to gain access to their legs. The waxy coating on their abdomen and legs was scored with an insect pin to provide better adhesion for the super glue. We made a pair of small holes a couple millimeters apart through the exoskeleton of their medial coxa on both the left mesothoracic and metathoracic legs to gain access to femoral extensor muscles 137 and 179 respectively. We then inserted insulated silver wire electrodes (0.003 in. wire diameter, A-M Systems, Sequim, WA) into the holes just underneath the exoskeleton and glued them in place. A fifth ground electrode was inserted and glued into the abdomen following the same procedure. The wires were routed along the abdomen and glued on one rostral and one caudal segment. The light tether trailed behind the cockroach and was elevated to a connector above the experimental chamber. These methods are similar to those in (4, 5).

Each electrode pair was amplified 100x using a differential amplifier (A-M Systems, Sequim, WA). Amplified signals were recorded through a data acquisition board (National Instruments, Austin, TX) and logged using custom software written in Matlab (Mathworks, Natick, MA). High speed video (Photron, San Diego, CA) was recorded at 500 fps from above. The arena was lighted with an array of infrared LEDs (Larson Electronics LLC, Kemp, TX). We prodded the cockroaches to run through a narrow opening that led to a wider field and recorded only trials in which the cockroach remained at least a centimeter from the walls. After 12 successful trials each lasting less than 2 seconds, videos were downloaded from the camera to a hard drive, and the cockroach rested for around 10 minutes until the next set of 12 trials. Up to 8 sets of 12 trials were collected per individual.

EMG data were processed offline using a digital bandpass filter. A simple peak finding method was used to discriminate spikes from the filtered EMG data. The 2D kinematics of the endpoints of all six legs were tracked semi-automatically in the horizontal plane from the high speed video using custom software written in Matlab. Cubic spline interpolation was used to estimate the position of the leg endpoints during occlusions. Interpolated kinematics were manually checked for a subset of videos to insure accuracy. A global phase variable was estimated using the Phaser algorithm (6) and subsequently used to separate both the EMG and leg kinematic data into individual strides. Stride frequency was estimated from the average change in global phase versus change in time over a stride.

**Dimensionality Reduction of Cockroach Output States*.** The output states for the cockroach, as well as all time series data, is too high dimensional to be able to effectively estimate mutual information. As the data is auto-correlated with time, we expect that a low dimensional representation of the states will contain all mutual information. The simplest dimensionality reduction is to take one sample from the trajectory in phase for each stride, which we call a phase slice. We found that adding a second slice increased the estimated information significantly, but not a third or fourth. The phases of the two slices also can result in various estimations of information as shown in Fig. S3. We thus chose two slices that were a half cycle apart that rested on the plateau of both the local and global MI landscape as shown by the black point in Fig. S3. We confirmed that conclusions concerning centralization did not change with as the phase of these slices varied or more slices were added.

We also tried other dimensionality reduction methods such as principle component analysis (7) and partial least squares (8). We found that the two phase slice method resulted in higher estimates of mutual information than the first two components of these other methods, although overall conclusions were robust to the different methods of dimensionality reduction.

**References**

1. Burks AW, Shannon CE, Weaver W (1951) The Mathematical Theory of Communication. *The Philosophical Review* 60(3):398.
2. Kraskov A, Stögbauer H, Grassberger P (2004) Estimating mutual information. *Physical Review E* 69(6):066138.
3. Srivastava KH, et al. (2017) Motor control by precisely timed spike patterns. *Proceedings of the National Academy of Sciences of the United States of America* 114(5):1171–1176.
4. Watson JT, Ritzmann RE (1998) Leg kinematics and muscle activity during treadmill running in the cockroach, Blaberus discoidalis: I. Slow running. *Journal of Comparative Physiology - A Sensory, Neural, and Behavioral Physiology* 182(1):11–22.
5. Sponberg S, Full RJ (2008) Neuromechanical response of musculo-skeletal structures in cockroaches during rapid running on rough terrain. *The Journal of experimental biology* 211(Pt 3):433–46.
6. Revzen S, Guckenheimer JM (2008) Estimating the phase of synchronized oscillators. *Physical Review E* 78(5):051907.
7. Pearson K (1901) Principal components analysis. *The London, Edinburgh, and Dublin Philosophical Magazine and Journal of Science* 6(2):559.
8. Sponberg S, Daniel TL, Fairhall AL (2015) Dual dimensionality reduction reveals independent encoding of motor features in a muscle synergy for insect flight control. *PLoS Computational Biology* 11(4):e1004168.

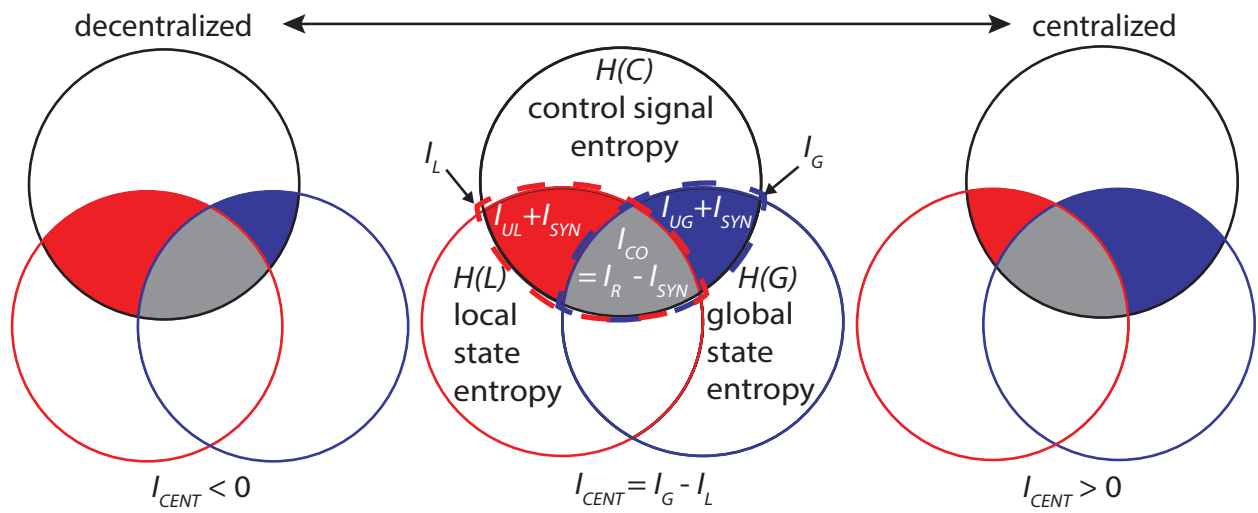

**Fig. S1.** Representation of overlapping entropies of the control signal, local state, and global state. We calculate $I_L$, the area encapsulated by the dotted red line, $I_G$, the area encapsulated by the dotted blue line, and $I_{TOT}$, which is the filled in red, blue, and gray areas. $I_{CENT}$ is negative when there is more red than blue area, and positive when there is more blue than red area.

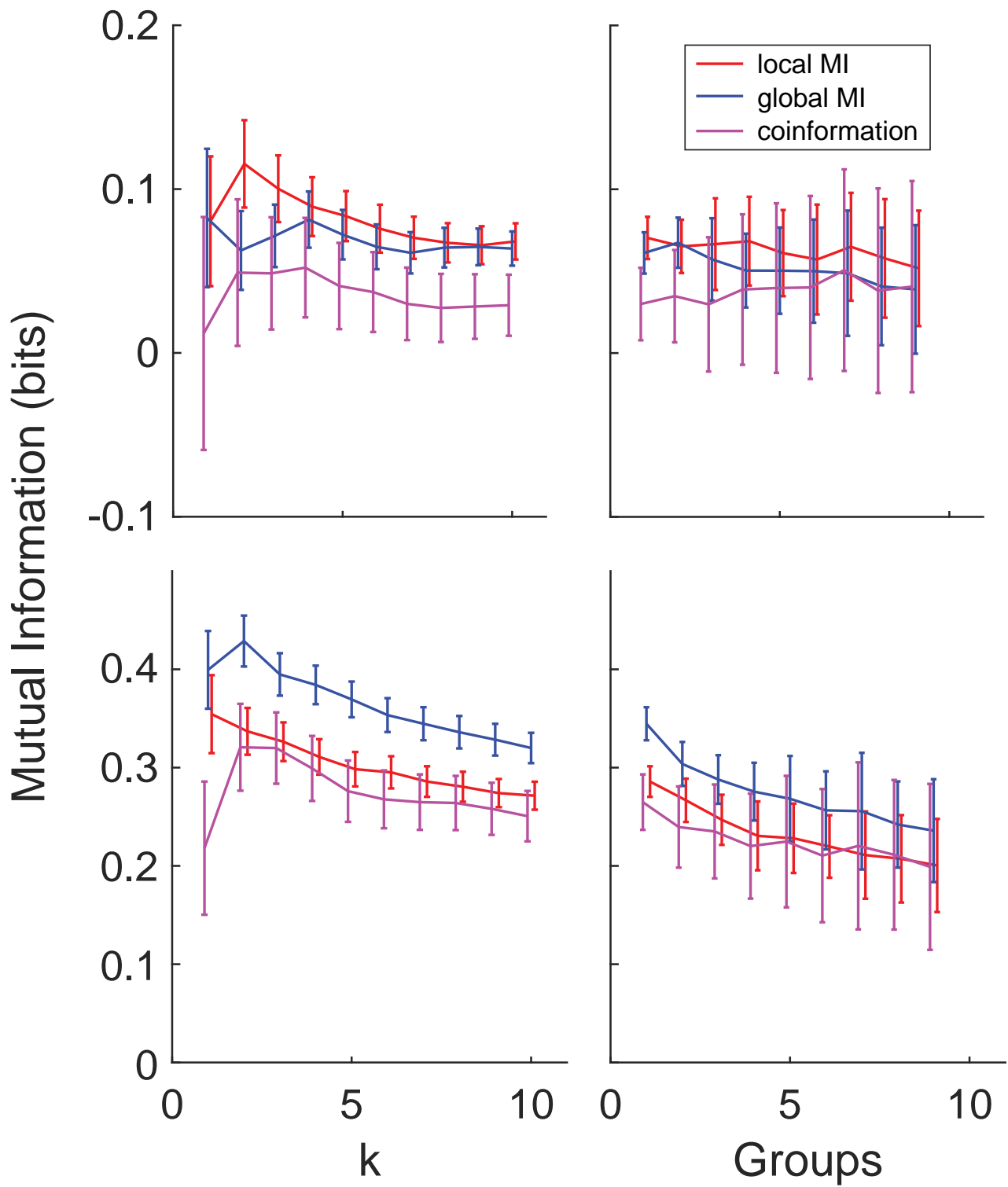

**Fig. S2.** Estimates of $I_G$ and $I_L$ while varying $k$ and sample size. Top row is information in count, bottom row is information in timing. The left column shows how the estimates vary with k using all data. The right column shows how estimates vary with the number of subdivided groups. Errorbars show the standard deviation of the estimate as calculated by the procedure given in the text and adapted from (3).

  Izaak D. Neveln, Amoolya Tirumalai and Simon Sponberg

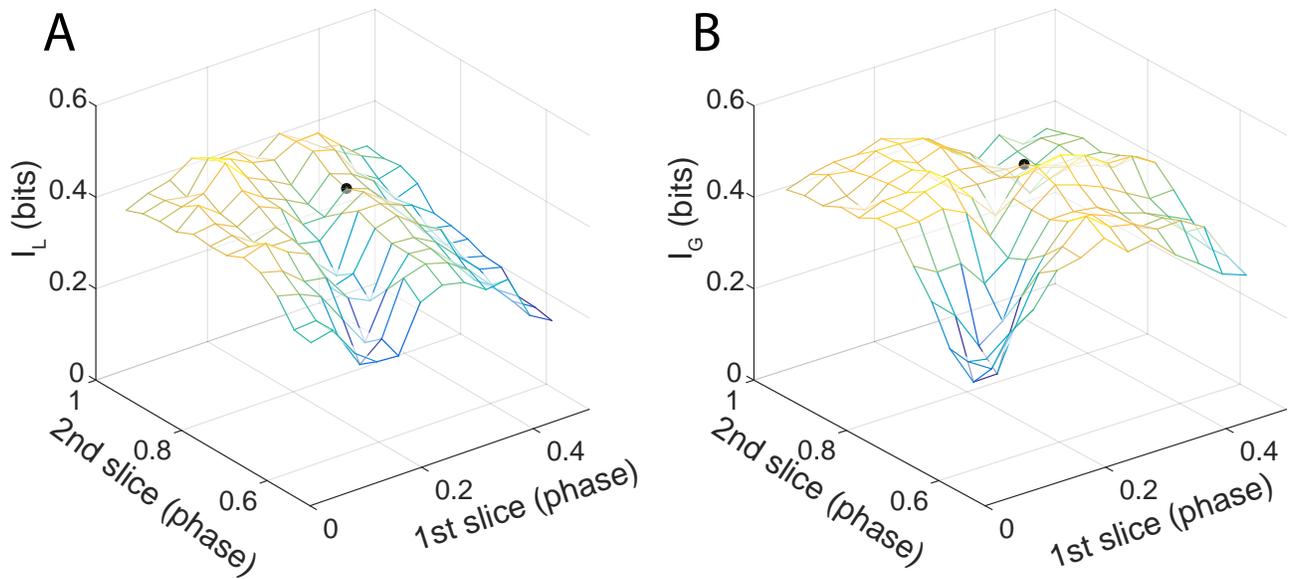

**Fig. S3.** Effect of phase slicing on mutual information estimates in the cockroach. A) $I_L$ (including both count and timing) estimates depend on which two slices of the local state are considered. B) Same as (A) for the $I_G$. We looked for a slice pair offset by a half cycle that resulted in a estimate close to maximal for both $I_L$ and $I_G$. We selected the two slices indicated by the black point and verified that conclusions concerning centralization did not change with small variations to the phase of these slices.